\documentstyle[12pt,epsfig]{article}
\def\be{\begin{equation}}
\def\ee{\end{equation}}
\def\bq{\begin{eqnarray}}
\def\eq{\end{eqnarray}}
\def\n{\nonumber}

\begin{document}

\begin{flushright}
SPhT t96/148 \\
December 1996
\end{flushright}
\vspace{1cm}
\begin{center}
{\bf A Note on the QCD Vacuum Tensor Susceptibility.}
\end{center}
\begin{center}{V.M. Belyaev$^*$ }
\end{center}
 \begin{center}
 {\em SPhT, CEA-SACLAY, 91191 Gif-sur-Yvette, CEDEX, France}
\end{center}
\begin{center}{A. Oganesian}\\
{\em ITEP, 117259, Moscow, Russia
}
\end{center} 
\vspace{1cm}
\begin{abstract}
We consider the procedure of determination of induced
quark condensates in the QCD sum rule approach.
It is noted that the previous estimations of this
value were very rough.
It is shown, that the detailed analyses of this parameter
of QCD vacuum leads to the conclusion that the value of the tensor
susceptibility has opposite sign and
much larger than estimations which were obtained before.
The reason of this contradiction is discussed.
It is noted that 
naive ways of determination of induced condensates
may lead to wrong results. The case of tensor susceptibility is the 
example  demonstrating the importance of the procedure which is applied
for calculation of such condensates.
New results for the nucleon tensor charge are presented.
The tensor charge is related to the first moment
of the  transversity distribution
function $h_1(x)$.

\vspace{0.5cm}

\noindent PACS number(s): 11.15.Tk, 11.55.Hx,  12.38.Lg, 13.60.Hb
\end{abstract}

\vspace{1cm}
\flushbottom{$^*\overline{On\;  leave\;  of\;  absence }
\;from\;  ITEP,\;  117259\;  Moscow,\; Russia.$}

 \newpage
\section{Introduction}

The QCD sum rule approach suggested by
Shifman, Vainshtein and Zakharov \cite{SVZ} is  a useful tool
to study  various properties of hadrons (see, for example \cite{rev}).
Ioffe suggested to use this approach
to study properties of baryons \cite{Ioffe}. 
This method has been applied for
calculations of
 magnetic moments \cite{is,by} and axial charge \cite{bk} of baryons.
In order to determine static properties of hadrons it was suggested to 
consider correlators of quark currents in the
presence of external constant classical field where nonperturbative
effects are taken into account in the so-called induced condensates
\cite{is} or (which is the same) in bilocal operators \cite{by}.
This approach was extended to the case of variable external
 field \cite{bk2}.

Recently, this version of QCD sum rules  has been
used for the evaluation of nucleon tensor charge \cite{ji1,ji2}.
The value of nucleon tensor charge is related to 
the first moment of the  transversity distribution
function $h_1(x)$ \cite{jj},
where $h_1(x)$ is an additional twist-two chirality violating 
structure function which can be measured in the Drell-Yan process 
with both beam and target transversely polarized \cite{h1}. 

To make a reasonable estimation for nucleon tensor charge, it
is necessary to evaluate nonperturbative effects which are related to induced
condensates.
The so-called tensor susceptibility corresponds to the lowest dimension
nonvanishing operator   and
may play an important role in QCD sum rule analyse.
The first estimations for this tensor susceptibility were made
in \cite{ji1}, where the authors claimed that this value is small
and does not effect on their result for nucleon tensor charge.

Here, we present new calculations of the tensor susceptibility and reanalyse the QCD
sum rules for tensor charge obtained in \cite{ji1,ji2}.
It is shown that this susceptibility is  several times larger and
has opposite sign than the first estimation.

The first results for the structure function $h_1(x)$ in QCD sum 
rule approach has been made by
Ioffe and Khodjamirian \cite{ik}. They used QCD sum rules for
four-point correlators which were in \cite{dis}
to study structure functions of deep-inelastic scattering.

\section{Tensor Susceptibility}

Let us consider the correlator of two tensor currents:
\bq
\int d^4x e^{ipx}<0|T\{\bar{q}\sigma_{\mu\nu}q(x),\bar{q}\sigma_{\alpha\beta}q(0)\}
|0>
\n
\\
=(g_{\mu\alpha}g_{\nu\lambda}-g_{\mu\beta}g_{\nu\alpha})\Pi_1(p^2)
\n
\\
+(g_{\mu\alpha}p_\nu p_\beta-g_{\nu\alpha}p_\mu p_\beta-g_{\mu\beta}p_\nu p_\alpha
+g_{\nu\beta} p_\nu p_\alpha)\Pi_2(p^2).
\label{1}
\eq
The tensor susceptibility is determined by $\Pi_1(0)$.

We cannot determine the value of the polarization operator for $p^2=0$ 
directly but
we can use the procedure
which was suggested in \cite{bk1} and consists in two steps:
1) Determination of  parameters of a model for the spectral density of the correlator;
2) Evaluation of the  correlator from a dispersion relation by using this
model  spectral density.
This technique has been applied to find the value of magnetic susceptibility of quark
condensate \cite{bk1} and other induced quark condensates in
the presence of an external vector field \cite{bk2}.

\subsection{QCD sum rule and spectral density}

Using OPE we found that
\bq
\Pi_1(Q^2)=\frac{1}{8\pi^2}Q^2\ln (Q^2/\Lambda^2)+
\frac1{24Q^2}<\frac{\alpha_s}{\pi}G^2>_0
\n
\\
-\frac{4}{9}
\left(\frac19-\frac13+\frac23+1\right)\frac{\alpha_s}{\pi}
\frac{(4\pi^2<\bar{\psi}\psi>_0)^2}{4 \pi^2 Q^4}+
O(Q^{-6}).
\label{2}
\eq
Here the coefficients $\left(\frac19-\frac13+\frac23+1\right)$
correspond to  contributions of the operators:
$(\bar\psi D_\alpha D_\beta D_\gamma\psi)$,
$(\bar\psi D_\alpha\psi G_{\beta\gamma})$,
$(\bar\psi\psi D_\alpha G_{\beta\gamma})$ and
$(\bar\psi\psi)^2$.

The correlator of two tensor currents has been considered in
\cite{reinders}. Note that the coefficient for the
last term in (\ref{2}) disagrees with the result obtained in \cite{reinders}.

We can formally write the following dispersion relation for
$\Pi_1(Q^2)$:
\bq
\Pi_1(Q^2)=\frac1\pi\int_0^\infty ds
\frac{Im \Pi_1(s)}{s+Q^2}.
\label{3}
\eq

This dispersion relation has ultraviolet divergence and
can be presented in the form of dispersional relation with subtractions.
It can be shown that the contribution of vector mesons with $J^{PC}=1^{--}$ to the
$\Pi_1(Q^2)$ is absent. Such mesons contribute to $\Pi_2(Q^2)$.
The  nonzero contribution into $\Pi_1(Q^2)$ comes from
the hadronic states with $J^{PC}=1^{+-}$ only.
The lowest state with these quantum numbers is $B_1(1231)$ resonance.  

Let us define
\bq
<0|\bar q\sigma_{\mu\nu}q|B_1>=i f_B\varepsilon_{\mu\nu\rho\lambda}q_\rho
e_\lambda.
\label{4}
\eq

Using the standard model for spectral density 
\bq
\frac{1}\pi Im\Pi_1(s)=f_B^2 m_B^2 \delta(s-m_B^2)+
\Theta(s-s_0)\frac{1}\pi Im \Pi^{pert.}(s),
\label{5}
\eq
we can find
the following
QCD sum rule for $f_B$:
\bq
\frac{M^4}{8\pi^2}
\left[
1-e^{-\frac{s_0}{M^2}}\left(1+\frac{s_0}{M^2}\right)
+\frac{\pi^2}{3}\frac{<\frac{\alpha_s}{\pi}G^2>_0}{M^4}
-\frac{52}{81}\frac{\alpha_s}{\pi}\frac{2(4\pi^2<\bar\psi\psi>_0)^2}{M^6}
\right]
\n
\\
=f_B^2m_B^2e^{-\frac{m_B^2}{M^2}}.
\label{6}
\eq
Here $m_B=1.23 GeV$ is the mass of $B_1$-meson, $s_0$ is the
continuum
 threshold, $\frac{\alpha_s}\pi\simeq 0.1$, $Im\Pi_1^{pert.}(s)$ is
the imaginary part of the perturbative contribution,
$<\frac{\alpha_s}{\pi}G^2>_0=0.012 GeV^4$, $4\pi^2|<\bar\psi\psi>_0|=
0.55 GeV^3$.

The best fit for QCD sum rule (\ref{6}) gives:  $8\pi^2 f_B^2=2.4-2.6 GeV^2$ with
$s_0\simeq 3 GeV^2$. Thus, we have determined  the parameters of the model spectral density (\ref{5}).

\subsection{Dispersion Relation}.

The tensor susceptibility is defined by the value of $\Pi_1(Q^2)$ for
$Q^2=0$.
As noted in the previous section, the dispersional relation (\ref{3})
has ultraviolet divergence. 
The nature of this divergence is clear:
the perturbative quark propagator in the presence of an external field
cannot be expanded near the point $x^2=0$. On other hand,
the perturbative part of this propagator is  considered separately
in the expression for quark propagator in the constant external tensor
field (see eq.(11) \cite{ji2}). So, to avoid  double counting, we
have to subtract the perturbative contribution from $\Pi_1(Q^2)$.
This procedure gives  the following result for the
nonperturbative part of polarization operator $\Pi^{np.}$:
\bq
\Pi_1^{np.}(Q)=\frac1\pi\int_0^\infty
\frac{Im \Pi_1(s)-Im \Pi_1^{pert.}(s)}{s+Q^2}ds.
\label{7}
\eq
For the value $\Pi_1^{np.}$ we can use a dispersion relation without
subtractions.
This procedure has been suggested in \cite{bk2} and gives a finite and 
selfconsistent result. 

In the case of tensor susceptibility the value for $\Pi^{np.}_1(0)$
is
\bq
\Pi_1^{np.}(0)=f_B^2-\frac{s_0}{8\pi^2}.
\label{8}
\eq

Using  the QCD sum rule for $f_B$ (\ref{6}) we can evaluate $\Pi_1^{np.}(0)$. The dependence of this parameter of the
nonperturbative propagator on $M^2$ for different $s_0$ is depicted
in Fig.1.
\begin{figure}
\begin{center}
\vspace*{-0.3cm}
\epsfig{file=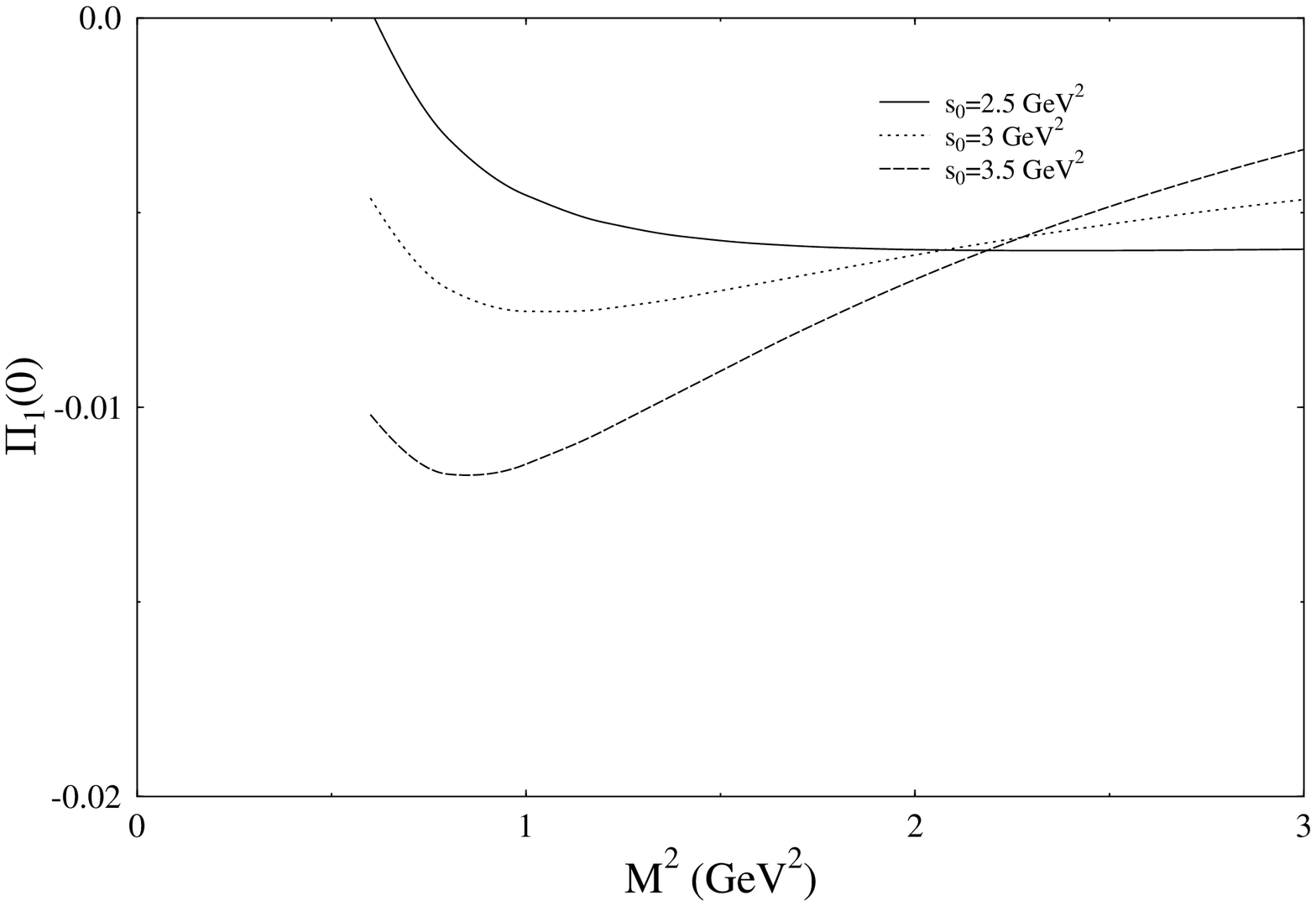,height=7.0cm}
\vspace*{-0.9cm}
\end{center}
\vspace*{-0.5cm}
\caption[]
 {}
\end{figure} 

From Fig.1 it is clear that 
\bq
\Pi_1^{np.}(0)\simeq -0.008 GeV^2.
\label{9}
\eq
The previous estimations of \cite{ji1,ji2} are:
\bq
\frac1{12}\Pi(0)=\Pi_1^{np.}\simeq 0.002 GeV^2.
\label{10}
\eq
Notice, that the result of present paper has opposite sign and is
 several times larger than the privious estimations \cite{ji1,ji2}.

\section{Discussion}

The result obtained in this
paper contradicts  the prediction 
which has been made in \cite{ji1,ji2}.
To understand what is a reason of such disagreement, let us consider
the dispersion relation for the polarization operator
\bq
\Pi(Q^2)=i\int d^4x e^{iqx}<0|T\{\bar q\sigma_{\mu\nu}q(x),
\bar q\sigma_{\mu\nu}q(0)\}|0>
\label{11}
\eq
which has been used in \cite{ji1,ji2}.
From eq.(\ref{1}) it is clear that
\bq
\Pi(0)=12\Pi_1(0).
\label{12}
\eq

In Ref.\cite{ji1} is was pointed that the perturbative contribution
to correlator (\ref{11}) is absent. 
It means that the dispersional relation for $\Pi(Q^2)$ 
gives a finite result for $\Pi(0)$ and we do not need to subtract
the perturbative contribution.
It was suggested to use $\rho(1^{--})$ and $B_1(1^{+-})$
dominance to estimate $\Pi(0)$.
It means that the dispersion relation for $\Pi(0)$ has the following form:
\bq
\Pi(0)=\frac1\pi\int\frac{\rho_\rho(s)+\rho_B(s)}sds
\label{13}
\eq
where only the  contributions of $\rho$- and $B_1$-mesons are taken into
accpount.
Perturbative contribution to the sum $\rho_\rho(s)+\rho_B(s)$
is absent. Naively we can expect  that  the meson dominance model
can give a reasonable prediction for $\Pi(0)$.
However, our results (\ref{9},\ref{10}) indicate that something is wrong.

To understand the source of discrepancy between two ways of the determinations
of the correlator, let us write  the following models for these
spectral densities:
\bq
\rho_\rho(s)=c_\rho \delta(s-m_\rho^2)+\Theta(s-s_\rho)\rho^{pert.}(s)
\label{14}
\\
\rho_B(s)=c_B\delta(s-m_{B_1}^2)-\Theta(s-s_B)\rho^{pert.}(s).
\label{15}
\eq
Here we use the fact that the perturbative contribution in the two channels
is equal to each other and has opposite sign.
It is important to note that it is possible to separate the contribution
$\rho$- and $B$-meson channels starting from correlator (\ref{1}).
The spectral densities of these two channels have nonvanishing
perturbative contribution and they are taking into account in
(\ref{14},\ref{15}).

Note that the thresholds $s_\rho$ and $s_B$ are not equal 
to each other.
It means, that there is additional (to $\rho$ and $B$-meson) contribution
in dispersion relation (\ref{13})
\bq
\int_{s_\rho}^{s_B}\frac{\rho^{pert.}(s)}{s}ds
\label{16}
\eq
which was not taken into consideration in \cite{ji1,ji2} and it is a source
of discrepancy between these results.
It is possible to reproduce our result by taking into account this
term.

Here we have presented the procedure which has to be used for correct evaluation
of induced condensates. Naive ways of determination of these condensates
may lead to wrong results. The case of tensor susceptibility is the good
example  demonstrating the importance of the procedure which is applied
for calculation of such condensates.

Finally we present our results for nucleon
tensor charge which can be obtained from QCD sum rules in ref.\cite{ji2}
using our result for the tensor susceptibility:
\bq
g_T^u=0.7\pm 0.3
\label{17}
\\
g_T^d\simeq 0.01.
\label{18}
\eq
These results were obtained from QCD sum rules for
$W_3$ (for $u$-tensor charge) and for $W_1$ (for $d$-tensor charge).
The previous results are: $g_T^u\simeq 1.33\pm 0.53$ and $g_T^d\simeq 0.04\pm 0.02$.
The tensor charge of nucleon is related to the first moment of  the
transversely distribution
function $h_1(x)$. It can be measured in the Drell-Yan process 
with both beam and target transversely polarized.

\section{Acknowledgements:}
We are very thankful to B.L. Ioffe for fruitful discussions
and to R. Peschanski for important remarks.
This work was supported in part by INTAS Grant 93-0283
and  by 
Swiss National Science Foundation.


\begin{thebibliography}{99}

\bibitem{SVZ}  M.A. Shifman, A.I. Vainshtein and V.I. Zakharov, Nucl. Phys.
{\bf B147} (1979) 385, 448.
\bibitem{rev} M.A. Shifman, (ed.), "Vacuum Structure and QCD Sum Rules",
Amsterdam, Netherlands: North-Holland (1992).
\bibitem{Ioffe} B.L. Ioffe, Nucl. Phys. {\bf B188} (1981) 317.
\bibitem{is} B.L. Ioffe and A.V. Smilga, JETP Lett. {\bf 37} (1983) 298;
Nucl. Phys. {\bf B232} (1984) 109.
\bibitem{by} Ya.Ya. Balitskii and A.V. Yung, Phys. Lett. {\bf B129} (1983) 328.
\bibitem{bk} V.M. Belyaev and Ya.I. Kogan, JETP Lett. {\bf 37} (1983) 730.
\bibitem{bk2} V.M. Belyaev and I.I. Kogan Int.J.Mod.Phys. {\bf A8} (1993) 153;
 preprint ITEP-12  (1984).
\bibitem{ji1} Hanxim He and Xiangdong Ji,
Phys. rev {\bf D52} (1995) 2960.
\bibitem{ji2} Hanxim He and Xiangdong Ji, Phys. Rev. {\bf D54} (1996) 6897.
\bibitem{jj} R.L. Jaffe and X. Ji, Phys. Rev. Lett. {\bf 67}
(1991) 552; Nucl. Phys. {\bf B375} (1992) 527.\
\bibitem{h1} J. Ralston and D.E. Soper, Nucl. Phys. {\bf B152}
(1979) 109.
\bibitem{ik} B.L. Ioffe and A. Khodjamirian, Phys. Rev. {\bf D51}
(1995) 3380.
\bibitem{dis} B.l. Ioffe, JETP Lett. {\bf 42} (1985) 327;
V.M. Belyaev and B.L. Ioffe, Nucl. Phys. {\bf B310} (1988) 548.
\bibitem{bk1} V.M. Belyaev and Ya.I. Kogan, Yad. Fiz. {\bf 40} (1984) 1035.
\bibitem{reinders} J. Govaerts, L.J. Reinders, F. De Viron and J. Weyers,
Nucl.Phys. {\bf B283} (1987) 706.

\end{thebibliography}
\end{document}